\documentclass{iaus}
\usepackage{graphicx,amsmath,natbib}

\title[Cold flows and mergers] %% give here short title %%
{The impact of ISM turbulence, clustered star formation and feedback on galaxy mass assembly through cold flows and mergers}

\author[Leila C. Powell et al]   %% give here short author list %%
{Leila C. Powell $^1$,
 Frederic Bournaud$^1$,
 Damien Chapon$^1$,
 Julien Devriendt$^2$,
 Adrianne Slyz$^2$
 \and Romain Teyssier$^{1}$}

\affiliation{$^1$ Service d'astrophysique, CEA, Orme des Merisiers, Gif-sur-Yvette Cedex, France  \\email: {\tt leila.powell@cea.fr}\\
$^2$ Oxford astrophysics, Denys Wilkinson Building, Keble Road, OX1 3RH \\
}%$^3$ Institute of Theoretical Physics, University of ZurichWinterhurerstrasse190, CH-8057 Zurich, Switzerland} 
\pubyear{2010}
\volume{277}  %% insert here IAU Symposium No.
\pagerange{--}
\jname{Tracing the ancetsry of galaxies}
\editors{Carignan, C., Freeman, K. C. \& Combes, F. eds.}
\begin{document}
\maketitle

\begin{abstract}

Two of the dominant channels for galaxy mass assembly are cold flows (cold gas supplied via the filaments of the cosmic web) and mergers. How these processes combine in a cosmological setting, at both low and high redshift, to produce the whole zoo of galaxies we observe is largely unknown. Indeed there is still much to understand about the detailed physics of each process in isolation. While these formation channels have been studied using hydrodynamical simulations, here we study their impact on gas properties and star formation (SF) with some of the first simulations that capture the multiphase, cloudy nature of the interstellar medium (ISM), by virtue of their high spatial resolution (and corresponding low temperature threshold). In this regime, we examine the competition between cold flows and a supernovae(SNe)-driven outflow in a very high-redshift galaxy ($z\approx9$) and study the evolution of equal-mass galaxy mergers at low and high redshift, focusing on the induced SF. We find that SNe-driven outflows cannot reduce the cold accretion at $z\approx 9$ and that SF is actually enhanced due to the ensuing metal enrichment. We demonstrate how several recent observational results on galaxy populations (e.g. enhanced HCN/CO ratios in ULIRGs, a separate Kennicutt Schmidt (KS) sequence for starbursts and the population of compact early type galaxies (ETGs) at high redshift) can be explained with mechanisms captured in galaxy merger simulations, provided that the multiphase nature of the ISM is resolved.

\keywords{methods: numerical, galaxies: evolution, galaxies: formation, galaxies: ISM, galaxies: star clusters }
\end{abstract}

\firstsection % if your document starts with a section,
\vspace{-0.4cm}
\section{The impact of feedback from individually-resolved SNe on a protogalaxy with a multiphase ISM}

The Nut simulations (presented in detail in \cite{me_2010}) probe the evolution of a protogalaxy embedded in the cosmic web at $z \approx 9$, at very high resolution (0.5pc physical in the densest regions) using the resimulation technique with the AMR code {\sc ramses} \citep{ramses}.The main simulation includes cooling, SF, SNe feedback (kinetic feedback modelled as Sedov blastwaves \citep{dubois_teyssier_sn}), metal enrichment and a UV background \citep{uvbackground}. A galactic wind develops, filling a large fraction of the volume out to $\approx6r_{\rm vir}$ with hot gas, which is punctured by cold filaments of the cosmic web that propagate all the way to the disc. SF rates are high for the galaxy halo mass ($1 {\rm M}_{\odot}$/yr for a $10^9 {\rm M}_{\odot}$ halo) and we report the formation of star-forming clumps with typical masses of $10^6 {\rm M}_{\odot}$. In order to isolate the impact of the SNe-driven galactic outflow, we undertake a parallel analysis of a control run with no SNe feedback. 

\subsection{Cold inflows versus hot outflows}

We establish temperature and density thresholds that separate out the filaments ($0.1 \le  \rho \le 10$ atoms ${\rm cm}^{-3}$, $T < 2 \times 10^4$K), the clumps  (i.e. condensed gas, such as the main disc and satellite galaxies) ($  \rho > 10$ ${\rm cm}^{-3}$, $T < 2 \times 10^4$K), cold diffuse gas ($ \rho < 0.1$ ${\rm cm}^{-3}$, $T < 2 \times 10^4$K), warm diffuse gas  ($ \rho < 10 $ ${\rm cm}^{-3}$, $2 \times 10^4$K $<T < 2 \times 10^5$K) and hot diffuse gas ($ \rho < 10$ ${\rm cm}^{-3}$, $T > 2 \times 10^5$K). We are then able to measure and compare the inflow and outflow rates in each component; we calculate the mass flux through spherical shells, centred on the halo centre, out to the virial radius. The full results of the inflow/outflow analysis are described in \cite{me_2010}, but here we recall the main conclusions:\\

\noindent \paragraph{\bf Inflows} The most striking result is that the far-reaching wind does not impede the accretion of gas onto the central galaxy. We find that the filaments are the dominant supply of gas to the disc, and since the gas in the filaments is dense and highly supersonic, it is not surprising that the wind has not been able to halt its progress. This suggests that for galaxies with host  haloes of this mass ($10^9 {\rm M}_{\odot}$) that are predominantly fed by cold streams, SNe-driven winds are unlikely to be able to limit accretion and therefore reduce SF in these systems, as is commonly suggested. \\

\noindent \paragraph{\bf Outflows} Note that we associate the galactic wind with the warm and hot diffuse components, defined above. We find that while the wind is far-reaching (extending to around $6 r_{\rm vir}$ by $z=9$), the mass outflow rate is only about $10$ per cent of the inflow rate; we do not find a massive wind as seen in low redshift observations \citep[e.g.][]{cmartin99_outflows}. We also report some outflow in the cold, diffuse component, which we attribute to gas that was in the region around the filaments being entrained by the wind and swept out; an effect that is seen in observed low-redshift winds.

\subsection{The resulting star formation}

  \begin{figure} 
  \centering
  \begin{minipage}{0.63\linewidth}
  \includegraphics[width=0.5\textwidth, trim=10mm 5mm 13mm 5mm, clip]{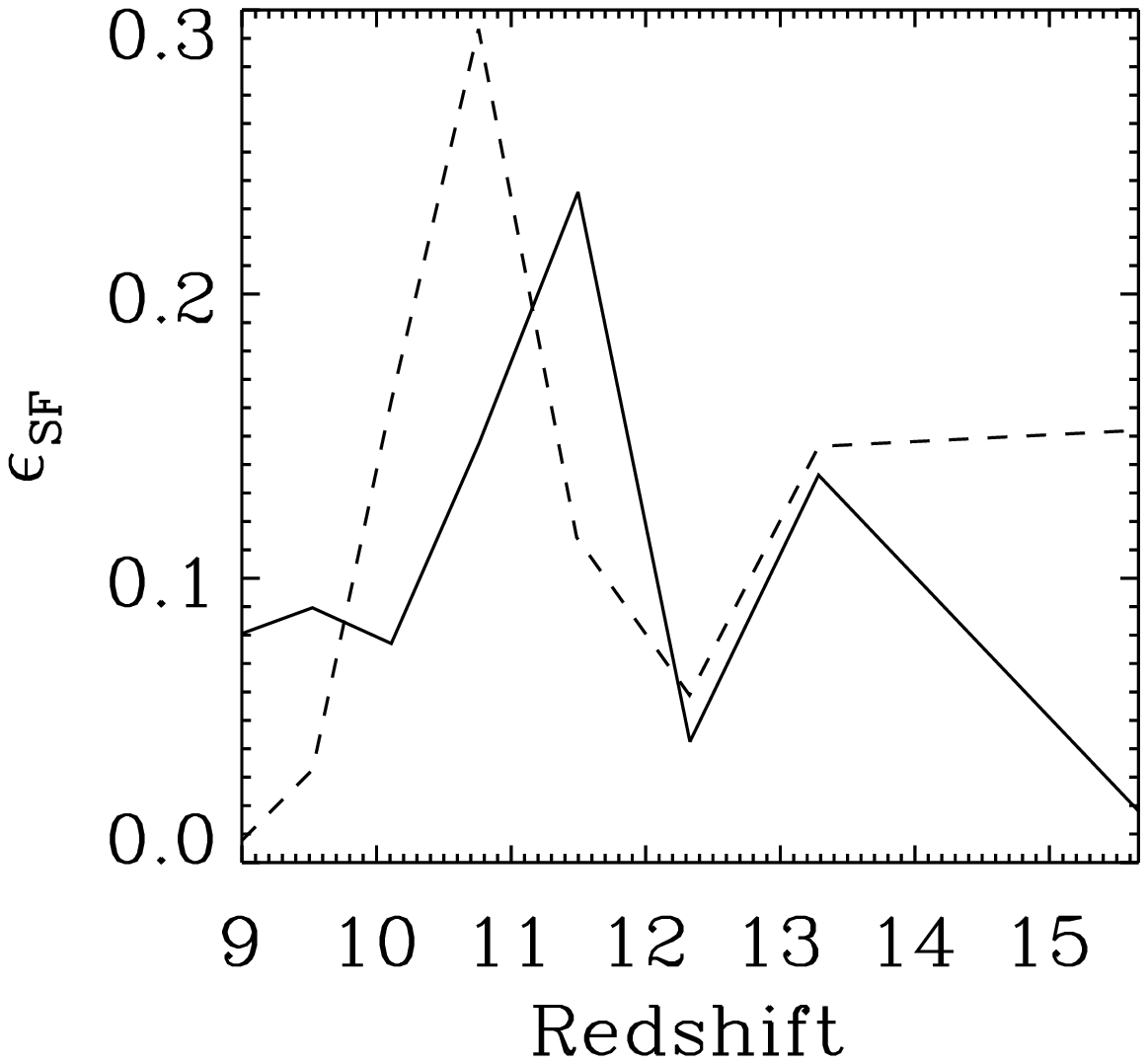}
   \includegraphics[width=0.5\textwidth, trim=10mm 5mm 13mm 3mm, clip]{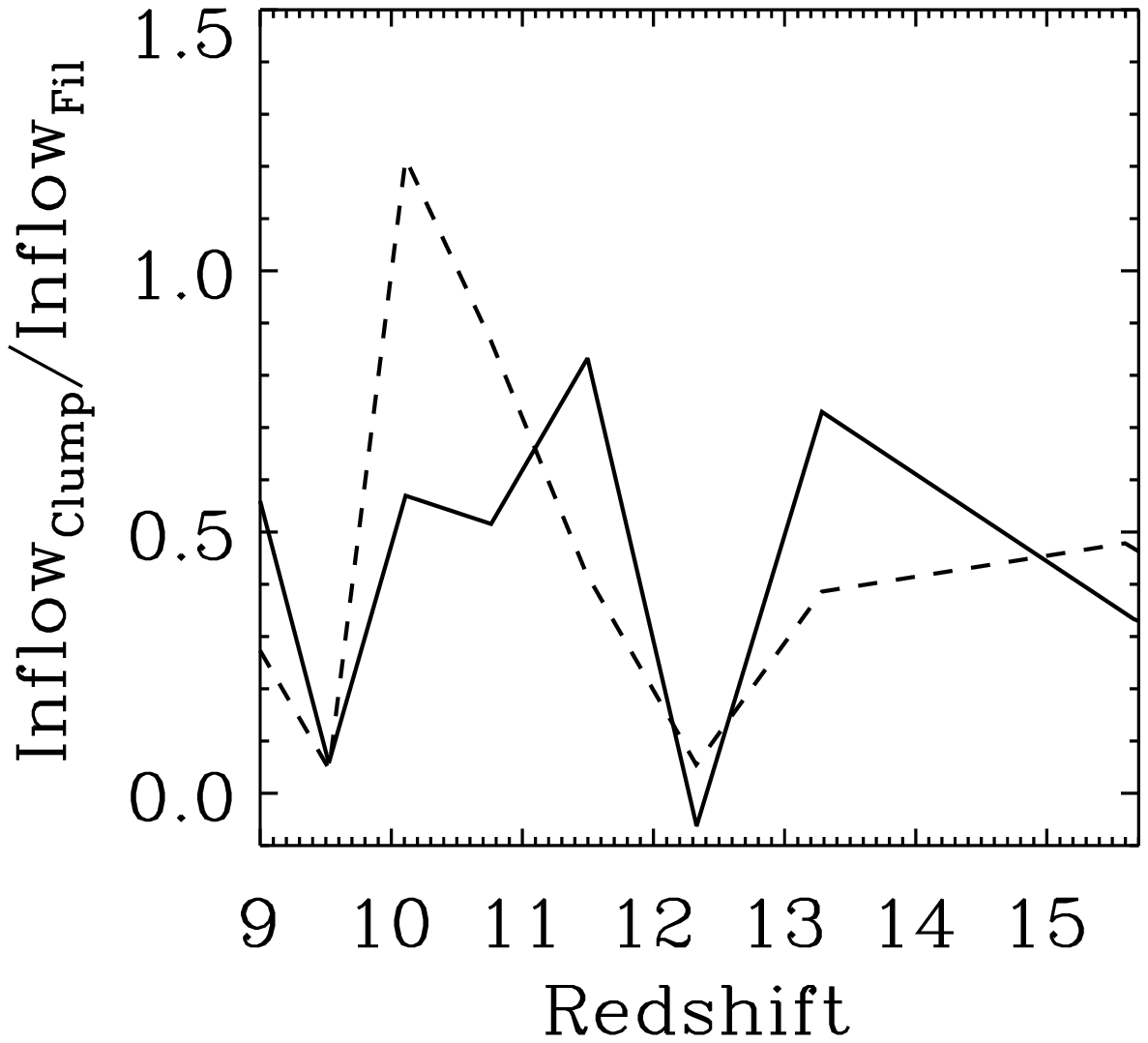}
   \end{minipage}
   \hspace{2mm}
   \begin{minipage}{0.34\linewidth}
   \centering
 \caption{{\bf Right: } The global average SF efficiency of the main galaxy versus redshift in the feedback (solid line) and no-feedback (dashed line) runs. {\bf Left: } Ratio of the net inflow in the clumpy component to the net inflow in the filamentary component versus redshift for the feedback (solid line) and no-feedback (dashed line) runs.}
    \label{fig1}
 \end{minipage}
\end{figure}

Surprisingly, we find the star formation rate (SFR) in the Nut simulation is a factor of  $\approx 10$  higher than that in the control run (without SNe) at $z=9$. We attribute this to the metal enrichment which results from the inclusion of SNe feedback. We find that the global average efficiency of SF is very high in both the Nut simulation and the control run. We use an efficiency of $1$ per cent to control SF on parsec-scales in {\sc ramses}, as found in the observations of \citet{krumholztan}. However the global efficiency is $\sim 10$ per cent as shown in Fig.~\ref{fig1} (left). We find that this efficiency is correlated with merger activity; in Fig.~\ref{fig1} (right) we show the ratio of inflow in the clumpy component to that in the filamentary component. High values of this quantity indicate that mergers are supplying a significant fraction of the gas being accreted; it is clear that these instances coincide with peaks in the global SF efficiency of the main galaxy (left).

These characteristics of the SF can be explained as follows. As evidenced by the star-forming clumps seen in the Nut simulation at $z=9$, and supported by clumps captured in other simulations \citep[e.g.][]{agertz_clumpygal} and observations of clump-cluster galaxies \citep[e.g.][]{clump_cluster}, all at $z=3$,  it seems probable that SF occurs in dense clumps at high redshift. We postulate that the correlation of the global SF efficiency with the relative importance of merger activity reflects the fact that mergers can induce fragmentation of the ISM, resulting in a more clumpy medium \citep{antennae} and more gas becoming eligible for SF. We examine the process of merger-induced SF in detail in the next section.

\vspace{-0.4cm}
\section{Merger-induced star formation at high resolution in a cloudy ISM}

Merger-induced clustered SF has been demonstrated in simulations of the Antennae system \citep{antennae}. We are undertaking a similar investigation, but with a whole sample of mergers with different orbital parameters, in order to assess the relative importance of this type of merger-induced SF. We have simulated 5 1:1 mergers, with AMR code {\sc ramses} with a maximum spatial resolution of $5$pc, allowing us to resolve densities up to $10^6 {\rm cm}^{-3}$. The sample is currently being extended both with additional orbits and different mass ratios. Here, we focus primarily on examining the mechanism via which SF is induced. 

\subsection{Changes in the star formation and ISM structure}

 \begin{figure} \centering
   \begin{minipage}{0.63\linewidth}
 \includegraphics[width=0.49\textwidth, trim=14mm 6mm 13mm 13mm, clip]{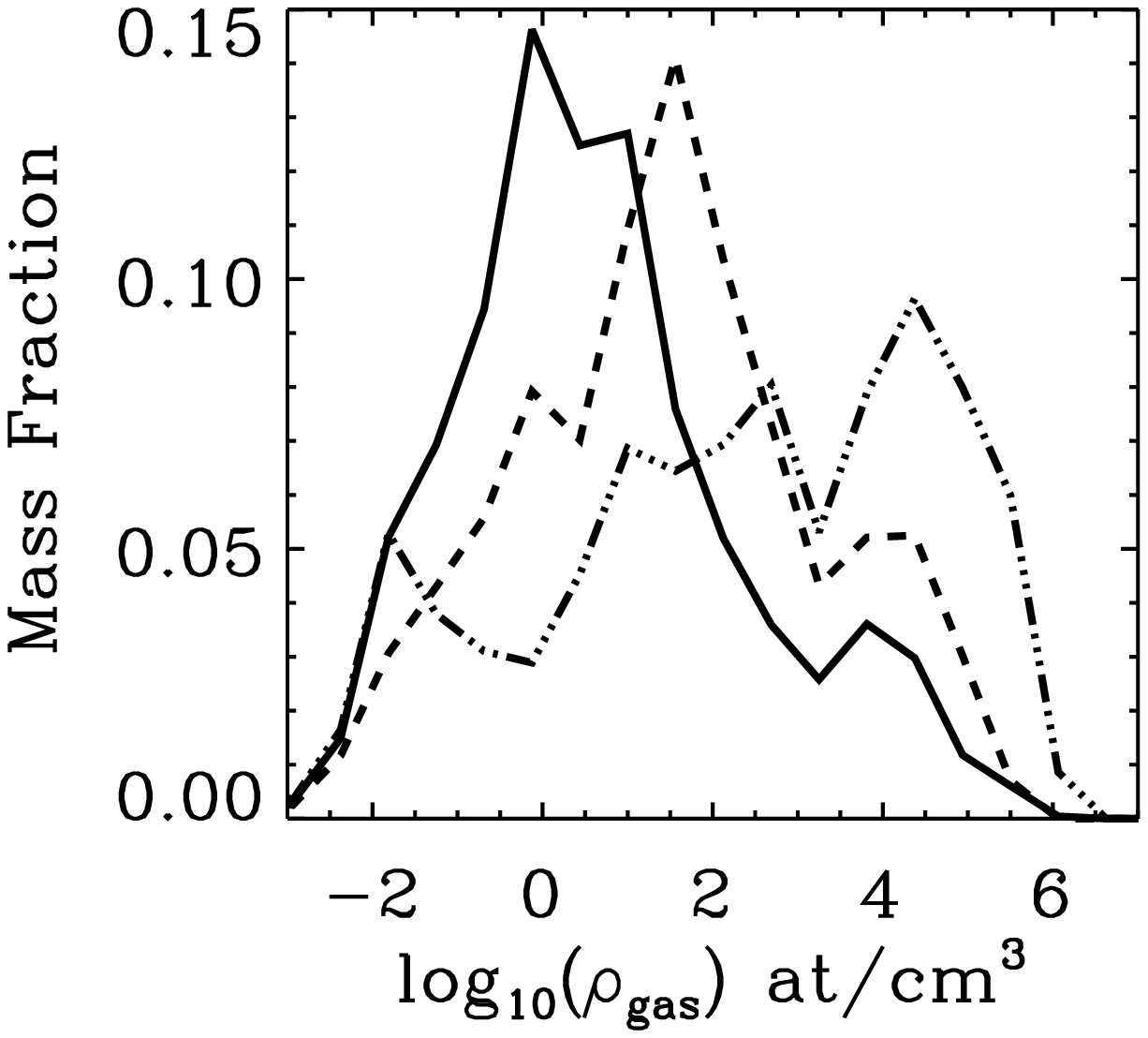}
  \includegraphics[width=0.5\textwidth, trim=14mm 6mm 11mm 11mm, clip]{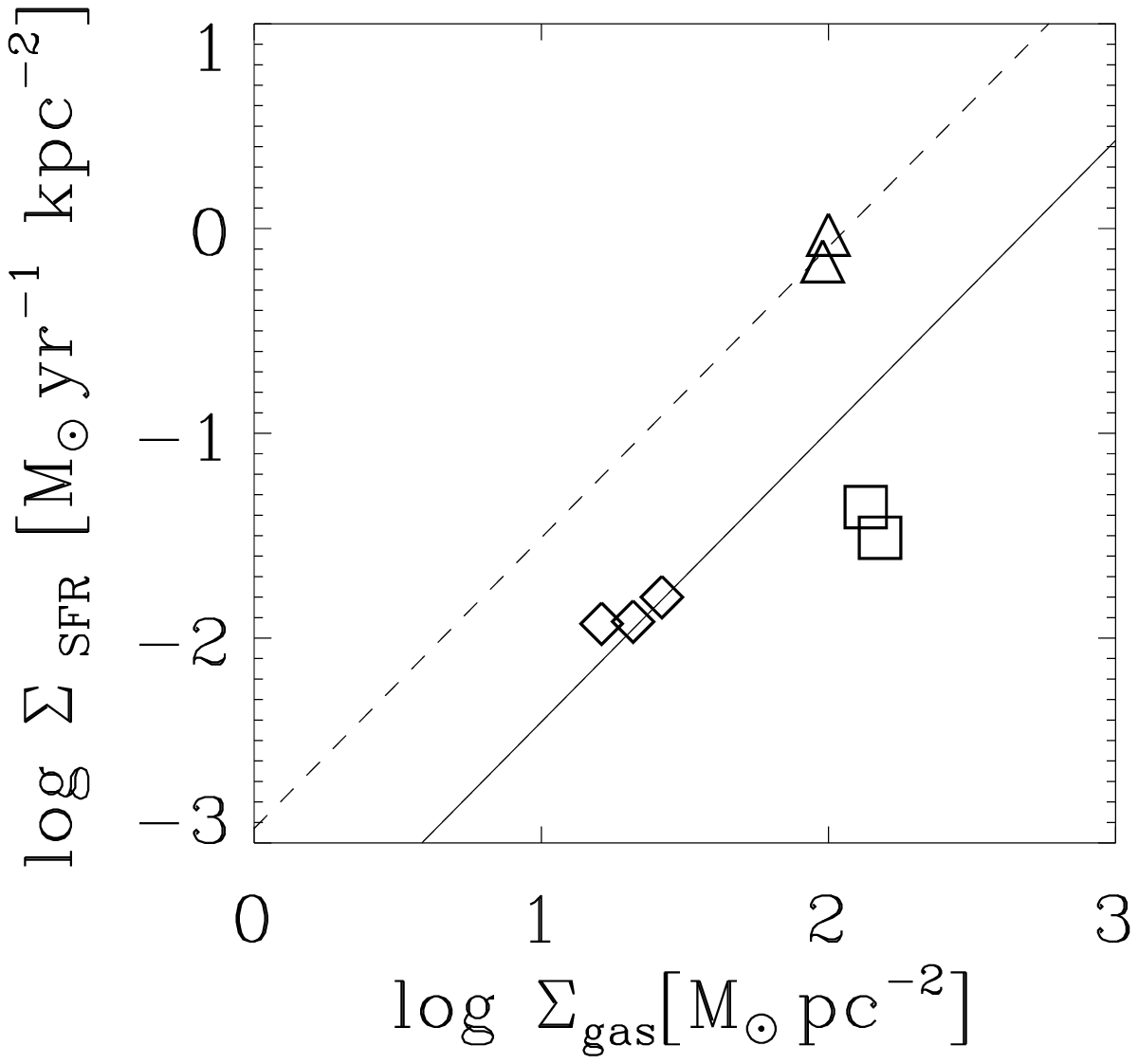}
  \end{minipage}
    \begin{minipage}{0.36\linewidth}
  \caption{{\bf Right: }Gas density pdf before (solid) and during (dashed) the merger and at the peak of the starburst (dot dash).  {\bf Left: }The KS relation for discs (solid) and starbursts (dashed) from \citet{daddi_etal_2010}, and selected points for simulated galaxies before the merger (diamonds), during the starburst (triangles) and after the merger (squares). In the latter stage the galaxy has been morphologically quenched \citep{MQ_2009}.}
  \label{fig2}
  \end{minipage}
  \end{figure}
    
We find that, in some cases, the peak in the SF rate occurs before the final coalesence of the two galaxies and the gas density distribution shows an extended, clumpy region, rather than a centrally-peaked but otherwise smooth distribution. In order to understand what impact the merger has had on the ISM we examine the gas density probability density function (pdf) in the left  of Fig.~\ref{fig2}. Here we can see the evolution of the pdf from before the merger (black line), during the initial stages of the merger (blue line) and at the peak of the starburst (red line). It is clear that there is an  excess of very dense gas ($\rho > 10^4$ ${\rm cm}^{-3}$) which increases as we progress through the merger. \\
\vspace{-0.5cm}
\subsection{Insights into observations of merging systems}

It is clear that in order to reproduce the formation of star clusters, which are observed both in high redshift clump-cluster galaxies and in lower redshift merging systems, it is necessary for hydrodynamical simulations to have a high spatial resolution (and the associated low temperature threshold). However, aside from the clustered SF, additional effects are apparent in such highly resolved simulations of merging systems which are not seen when the ISM is warm and smooth (as in low resolution simulations). \\

\begin{itemize}

\item{We have captured an excess of dense gas directly in our merger simulations (as demonstrated in the left  of Fig.~\ref{fig2}). It was proposed (and demonstrated using post-processing of simulations) by \citet{juneau_etal_2009} that an excess of dense gas in the ISM could be induced by mergers and that, since HCN emission traces denser gas phases than CO, this could account for observations of enhanced HCN/CO ratios in ULIRGs. This would remove the need to invoke active galactic nuclei activity.}\\

\item{The increased fraction of gas mass in this very dense phase (in the form of clumps) can also provide a physical explanation for the two KS relations demonstrated in \citet{daddi_etal_2010}.} Fig.~\ref{fig2} (right ) shows that as our merger simulations progress, the galaxies move up and to the right in the $\Sigma_{\rm SFR}$-$\Sigma_{\rm gas}$ plot, reaching the starburst sequence at the peak of the SFR, then moving down to below the disc sequence once much of the fuel has been consumed. There is some central gas inflow which increases the average gas surface density $\Sigma_{\rm gas}$. However, the merging galaxies reach the starburst sequence because $\Sigma_{\rm SFR}$ is much higher than would be expected from a smooth gas distribution with the same value of $\Sigma_{\rm gas}$ due to the high efficiency of SF in the very dense gas clumps.\\

\item{Simulations of high redshift mergers (which have a higher gas fraction than those discussed above) show that the resolving the clumpy nature of the ISM can also have a significant impact on the merger remnant \citep{bournaud_etal_2010}.} There are two main effects when the ISM is multiphase: 1) the merger remnant is much more compact and 2) there is a much smaller surviving/reformed disc component. The first conclusion provides a possible formation mechanism for the population of compact ETGs at high redshift, while the latter casts doubt on the likelihood of discs surviving/reforming after mergers without some external source of gas (e.g. cold accretion).

\end{itemize}

\vspace{-0.4cm}
\section{Conclusions}

We demonstrate that many insights into galaxy mass assembly via both cold flows and mergers can be gleaned from simulations with sufficient resolution to capture the cloudy nature of the ISM and clustered SF. We find that SNe-driven outflows cannot reduce the cold accretion onto dwarf-mass galaxies at $z\approx 9$ and that the SFR is actually higher due to the ensuing metal enrichment. We describe how high-resolution galaxy merger simulations enable us to explain several observational results (e.g. enhanced HCN/CO ratios in ULIRGs, a KS sequence for starbursts and high-redshift compact ETGs) as they reveal physical processes not captured previously.


\vspace{-0.4cm}
\begin{thebibliography}{}

\bibitem[\protect\citeauthoryear{{Agertz}, {Teyssier} \& {Moore}}{{Agertz}
  et~al.}{2009}]{agertz_clumpygal}
{Agertz} O.,  {Teyssier} R.,    {Moore} B.,  2009, MNRAS, 397, L64

\bibitem[\protect\citeauthoryear{{Bournaud}, F. and {Chapon}, D. and {Teyssier}, R. and {Powell}, L.~C. and {Elmegreen}, B.~G. and {Elmegreen}, D.~M. and {Duc}, {P.-A.} and {Contini}, T. and {Epinat}, B. and {Shapiro}, K.~L.}{Bournaud et al}{2010}]{bournaud_etal_2010}
{{Bournaud} F. et al}, 2010, arXiv:1006.4782

 \bibitem[\protect\citeauthoryear {{Daddi}, E. and {Elbaz}, D. and {Walter}, F. and {Bournaud}, F. and {Salmi}, F. and {Carilli}, C. and {Dannerbauer}, H. and {Dickinson}, M. and {Monaco}, P. and {Riechers}, D.}{Daddi et al.}{2010}]{daddi_etal_2010}
 {{Daddi}, E. et al}, 2010, ApJL, 714, 118

\bibitem[\protect\citeauthoryear{{Dubois} \& {Teyssier}}{{Dubois} \&
  {Teyssier}}{2008}]{dubois_teyssier_sn}
{Dubois} Y.,  {Teyssier} R.,  2008, A\&A, 477, 79


 \bibitem[\protect\citeauthoryear{{Elmegreen} and {Elmegreen}}{{Elmegreen} \& {Elmegreen}}{2005}]{clump_cluster}
 {{Elmegreen}, B.~G. \& {Elmegreen}, D.~M.}, 2005, ApJ, 627, 632
 
\bibitem[\protect\citeauthoryear{{Haardt} \& {Madau}}{{Haardt} \&
  {Madau}}{1996}]{uvbackground}
{Haardt} F.,  {Madau} P.,  1996, ApJ, 461, 20

 \bibitem[\protect\citeauthoryear {{Juneau}, S. and {Narayanan}, D.~T. and {Moustakas}, J. and {Shirley}, Y.~L. and {Bussmann}, R.~S. and {Kennicutt}, R.~C. and {Vanden Bout}, P.~A.}{Juneau et al}{2009}]{juneau_etal_2009}
{{Juneau}, S. et al}, 2009, ApJ, 707, 1217


\bibitem[\protect\citeauthoryear{{Krumholz} and {Tan}}{{Krumholz} and {Tan}}{2007}]{krumholztan}
   {{Krumholz}, M.~R. \& {Tan}, J.~C.}, 2007, ApJ, 654, 304


\bibitem[\protect\citeauthoryear{{Martig} et al.}{2009}]{MQ_2009}
{{Martig}, M. and {Bournaud}, F. and {Teyssier}, R. and {Dekel}}, A., 2009, ApJ, 707, 250

\bibitem[\protect\citeauthoryear{{Martin}, C.~L.}{Martin}{1999}]{cmartin99_outflows}
 {{Martin}, C.~L.}, 2009, ApJ, 513, 156

\bibitem[\protect\citeauthoryear{{Powell}, {Slyz}, A. and {Devriendt}}{Powell et al}{2010}]{me_2010}
{{Powell}, L.~C. , {Slyz}, A. \& {Devriendt}, J.},2010, arXiv:1012.2839

\bibitem[\protect\citeauthoryear{{Teyssier}}{{Teyssier}}{2002}]{ramses}
{Teyssier} R.,  2002, A\&A, 385, 337

\bibitem[\protect\citeauthoryear{{Teyssier}, {Chapon} \& {Bournaud}}{{Teyssier}
  et~al.}{2010}]{antennae}
{Teyssier} R.,  {Chapon} D. \&  {Bournaud} F.,  2010, ApJL, 720, 149


  \end{thebibliography}
\end{document}